\documentstyle[12pt]{article}
\newlength{\extraspace}
\newlength{\extraspaces}
\newcommand{\be}{\begin{equation}
\addtolength{\abovedisplayskip}{\extraspaces}
\addtolength{\belowdisplayskip}{\extraspaces}
\addtolength{\abovedisplayshortskip}{\extraspace}
\addtolength{\belowdisplayshortskip}{\extraspace}}
\newcommand{\ee}{\end{equation}}
\newcommand{\ba}{\begin{eqnarray}
\addtolength{\abovedisplayskip}{\extraspaces}
\addtolength{\belowdisplayskip}{\extraspaces}
\addtolength{\abovedisplayshortskip}{\extraspace}
\addtolength{\belowdisplayshortskip}{\extraspace}}
\newcommand{\ea}{\end{eqnarray}}
\newcommand{\newsection}[1]{
\vspace{7mm}
\pagebreak[3]
\addtocounter{section}{1}

{\large {\bf \thesection. #1}}
\nopagebreak
\medskip
\nopagebreak
\hspace{3mm}}
\newcommand{\nonu}{\nonumber \\[.5mm]}
\newcommand{\A}{&\!\!\!}

\begin{document}
\addtolength{\baselineskip}{3.0mm}

%
%

\vspace{3cm}

\begin{center}
{\large{\bf{Construction of 
            $N = 2$ chiral supergravity
            
            compatible with the reality condition}}} 
\\[20mm]
{\large Motomu Tsuda and Takeshi Shirafuji} \\[12mm]
Physics Department, Saitama University \\[2mm]
Urawa, Saitama 338, Japan \\[20mm]
{\bf Abstract}\\[10mm]
{\parbox{13cm}{\hspace{5mm}
We construct $N = 2$ chiral supergravity (SUGRA) 
which leads to Ashtekar's canonical formulation. 
The supersymmetry (SUSY) transformation 
parameters are not constrained at all 
and auxiliary fields are not required 
in contrast with the method of the two-form gravity. 
We also show that our formulation is compatible 
with the reality condition, 
and that its real section is reduced to the usual 
$N = 2$ SUGRA up to an imaginary boundary term. }} 
\end{center}
\vfill

\newpage
\newsection{Introduction}

To introduce a self-dual connection into general relativity 
has the advantage of leading to polynomial constraints 
in the canonical formulation and therefore of obtaining 
a nonperturbative quantum gravity \cite{AA,JS,RS,Ro}. 
However, the extension of Ashtekar's formulation 
of general relativity to include matter spinor fields 
is not trivial, because the Lagrangian using the complex 
self-dual connection may lead to an imaginary part which gives 
extra conditions or inconsistencies for field equations. 
Indeed its imaginary part for the real tetrad has 
a particular quadratic form of the torsion tensor, 
i.e., $\epsilon^{\mu \nu \rho \sigma} T_{\lambda \mu \nu} 
T{^{\lambda}}_{\rho \sigma}$ \cite{Do,TS1}. 
\footnote{\ 
We shall follow the notation and convention 
of Ref.\cite{TS1,TS2}.}
In the second-order formalism this term vanishes 
for spin-1/2 fields \cite{Do,TS1,JJ1,ART} 
because the torsion tensor is totally antisymmetric, 
and it vanishes also for only one spin-3/2 field 
\cite{TS1,JJ2} after a Fierz transformation. 
For two or more spin-3/2 fields, however, 
the quadratic term 
$\epsilon^{\mu \nu \rho \sigma} T_{\lambda \mu \nu} 
T{^{\lambda}}_{\rho \sigma}$ does not vanish 
even if Fierz transformations are used. 
If this imaginary term is not canceled with another 
appropriate term in the Lagrangian, then it will 
give additional equations for spin-3/2 fields 
which cause inconsistency \cite{TS1}. 
Therefore the consistent inclusion of $N = 2$ 
supergravity (SUGRA) into Ashtekar's formulation, 
i.e., the construction of $N = 2$ chiral SUGRA is not 
a trivial problem. 

One method to construct the chiral SUGRA is based on 
the two-form gravity \cite{CDJ}, 
in which the ``chiral" means that 
the formulation is given by using only those quantities 
with unprimed spinor indices in the 2-component spinor 
notation. The $N = 1$ two-form SUGRA was first formulated 
by Capovilla et al.\cite{CDJ} in first-order form, 
and its extension to $N = 2$ SUGRA has also been 
established \cite{KS}. In this formulation, however, 
supersymmetry (SUSY) transformation parameters are 
constrained, and auxiliary fields are needed to formulate 
the chiral theory in the above sense. 

On the other hand, we have reformulated \cite{TS2} 
Jacobson's construction of $N = 1$ chiral SUGRA \cite{JJ2} 
into first-order form, closely following the method 
originally employed in the usual $N = 1$ SUGRA 
\cite{FNF,DZ}. 
\footnote{\ 
The minimal off-shell version of $N = 1$ chiral SUGRA 
can also be constructed by introducing an antisymmetric 
tensor field and an axial-vector field 
as auxiliary fields \cite{TS3}.}
In this formulation the meaning of the ``chiral" is 
weaker than in the two-form gravity: 
Namely, both primed and un-primed spinor indices 
are used in the 2-component spinor notation, 
although only right-handed (or left-handed) spin-3/2 
fields are coupled to the spin connection. 
The merit of this method lies in that SUSY transformation 
parameters are not constrained at all, and that 
auxiliary fields are not required. 
Therefore it is suitable for discussing its relation 
to the usual SUGRA and, in particular, for studying 
the consistency problem which may arise when two or more 
spin-3/2 fields are included. 

In this paper we construct $N = 2$ chiral SUGRA 
using the SUSY transformation parameters without constraints. 
We also show that the dangerous term quadratic in torsion, 
$\epsilon^{\mu \nu \rho \sigma} T_{\lambda \mu \nu} 
T{^{\lambda}}_{\rho \sigma}$, exactly cancels with 
another term which is required for the SUSY invariance, 
and therefore that the consistency problem 
does not arise even if the tetrad is real. 
Our construction closely follows the usual $N = 2$ SUGRA 
\cite{FN}: Namely, we try to couple the spin-(2, 3/2) 
chiral action \cite{TS2,JJ2} to the spin-(3/2, 1) 
supermatter multiplet by means of the Noether method 
\cite{FN,FGSN,West,Nieu}. 

In order to discuss the consistency problem, i.e., 
the question of whether a real section can be extracted 
>from the theory or not, we assume at first 
that the tetrad $e^i_{\mu}$ is complex, 
and we introduce two independent pairs of (Majorana) 
Rarita-Schwinger fields, $(\psi_{\mu}, \tilde \psi_{\mu})$ 
and $(\varphi_{\mu}, \tilde \varphi_{\mu})$, 
\footnote{\ 
We assume $\psi_{\mu}$ and $\tilde \psi_{\mu}$ 
(or $\varphi_{\mu}$ and $\tilde \varphi_{\mu}$) 
to be two independent (Majorana) Rarita-Schwinger fields, 
and define the right-handed spinor fields 
$\psi_{R \mu} := (1/2)(1 + \gamma_5) \psi_{\mu}$ 
and $\tilde \psi_{R \mu}:= (1/2)(1 + \gamma_5) 
\tilde \psi_{\mu}$. 
The $\psi_{R \mu}$ and $\tilde \psi_{R \mu}$ 
are related to the left-handed spinor fields 
$\psi_{L \mu}$ and $\tilde \psi_{L \mu}$ respectively, 
because $\psi_{\mu}$ and $\tilde \psi_{\mu}$ 
are Majorana spinors.
We denote the Dirac conjugate of $\psi_{R \mu}$ 
by $\overline \psi_{L \mu}$.}
along with a complex spin-1 field $A_{\mu}$. 
This means that right- and left-handed SUSY 
transformations introduced in the chiral SUGRA are 
independent of each other even in the second-order formalism. 
This fact makes it more transparent to confirm the SUSY 
invariance, particularly the right-handed one, 
in the course of constructing the theory. 
At the final stage we impose the reality condition. 

This paper is organized as follows. In section 2 we summarize 
$N = 1$ chiral SUGRA in first-order form. 
In section 3 we construct $N = 2$ chiral SUGRA. 
A manifestly $O(2)$ invariant form 
of $N = 2$ chiral Lagrangian is explicitly shown in section 4. 
In section 5 we discuss the consistency problem 
and the relation to the usual $N = 2$ SUGRA 
within the second-order formalism. 
Finally, the conclusion and discussion are given in section 6.

\newsection{Summary of $N = 1$ chiral SUGRA}

Let us first summarize the chiral Lagrangian 
of $N = 1$ theory and its local right- and left-handed 
SUSY transformation laws in first-order form \cite{TS2}. 
The independent variables are the complex tetrad 
$e^i_{\mu}$, two independent (Majorana) Rarita-Schwinger 
fields $(\psi_{\mu}, \tilde \psi_{\mu})$ 
and the self-dual connection $A^{(+)}_{ij \mu}$ 
which satisfies 
$(1/2){\epsilon_{ij}} \! ^{kl} A^{(+)}_{kl \mu} 
= i A^{(+)}_{ij \mu}$. The chiral Lagrangian density 
is given by 
\footnote{\ 
The $\kappa^2$ is the Einstein constant: 
$\kappa^2 = 8 \pi G/c^4$. We take the units 
$c = \hbar = 1$.}
\be
{\cal L}^{(+)}_{SG} = - {i \over {2 \kappa^2}} 
                      e \ \epsilon^{\mu \nu \rho \sigma} 
                      e^i_{\mu} e^j_{\nu} 
                      R^{(+)}_{ij \rho \sigma} 
                    - e \ \epsilon^{\mu \nu \rho \sigma} 
                      \overline{\tilde \psi}_{R \mu} 
                      \gamma_\rho D^{(+)}_\sigma \psi_{R \nu},
\label{LSG+}
\ee
where $e$ denotes ${\rm det}(e^i_{\mu})$, 
and the covariant derivative $D^{(+)}_\mu$ and 
the curvature ${R^{(+)ij}}_{\mu \nu}$ are 
\ba
\A \A D^{(+)}_\mu := \partial_\mu + {i \over 2} A^{(+)}_{ij \mu} 
      S^{ij}, \nonu
\A \A {R^{(+)ij}}_{\mu \nu} := 2(\partial_{[\mu} {A^{(+)ij}}_{\nu]} 
             + {A^{(+)i}}_{k [\mu} {A^{(+)kj}}_{\nu]}). 
\ea

The spin-(2, 3/2) Lagrangian density of (\ref{LSG+}) 
is invariant under the right-handed SUSY transformations 
generated by a Majorana spinor parameter $\alpha$, 
\ba
\A \A \delta_R e^i_{\mu} 
      = i \kappa \ \overline \alpha_L \gamma^i 
      \tilde \psi_{L \mu}, \nonu
\A \A \delta_R \psi_{R \mu} 
      = {2 \over \kappa} D^{(+)}_{\mu} \alpha_R, 
      \ \ \ \ \ \delta_R \tilde \psi_{L \mu} = 0, \nonu
\A \A \delta_R A^{(+)}_{ij \mu} = 0, 
\label{RSUSY-SG}
\ea
and also under the left-handed SUSY transformations 
generated by another Majorana spinor parameter 
$\tilde \alpha$, 
\footnote{\ 
We define the self-dual and antiself-dual part of a tensor 
$F_{ij}$ as $F^{(\pm)}_{ij} = {1 \over 2} (F_{ij} \mp 
{i \over 2} \epsilon_{ijkl} F^{kl})$.}
\ba
\A \A \delta_L e^i_{\mu} 
      = i \kappa \ \overline{\tilde \alpha}_R \gamma^i 
      \psi_{R \mu}, \nonu
\A \A \delta_L \psi_{R \mu} = 0, 
      \ \ \ \ \ \delta_L \tilde \psi_{L \mu} 
      = {2 \over \kappa} D^{(-)}_{\mu} \tilde \alpha_L, \nonu
\A \A \delta_L A^{(+)}_{ij \mu} 
      = {\rm self\! \! -\! \! dual\ part\ of\ } \{ - \kappa 
      (B_{\mu ij} - e_{\mu [i} B{^m}_{\mid m \mid j]}) \}
\label{LSUSY-SG}
\ea
with 
\be
B^{\lambda \mu \nu} 
:= \epsilon^{\mu \nu \rho \sigma} 
   \overline{\tilde \alpha}_R \gamma^{\lambda} 
   D^{(+)}_{\rho} \psi_{R \sigma}. 
\ee
Here the antiself-dual connection $A^{(-)}_{ij \mu}$, 
which appears in the left-handed SUSY transformations 
(\ref{LSUSY-SG}) through the covariant derivative 
$D^{(-)}_{\mu}$, is defined to be the solution 
of the equation 
$\delta {\cal L}^{(-)}_{SG}/\delta A^{(-)} = 0$ 
for the ``unphysical'' Lagrangian density 
\be
{\cal L}^{(-)}_{SG} = {i \over {2 \kappa^2}} 
                      e \ \epsilon^{\mu \nu \rho \sigma} 
                      e^i_{\mu} e^j_{\nu} 
                      R^{(-)}_{ij \rho \sigma} 
                    + e \ \epsilon^{\mu \nu \rho \sigma} 
                      \overline \psi_{L \mu} \gamma_\rho 
                      D^{(-)}_\sigma \tilde \psi_{L \nu}, 
\label{LSG-}
\ee
which becomes just the complex conjugate of 
${\cal L}^{(+)}_{SG}$ when the tetrad is real and 
$\tilde \psi_{\mu} = \psi_{\mu}$. 
On the other hand, we take the self-dual connection $A^{(+)}_{ij \mu}$ 
as one of the independent variables.

\newsection{Construction of $N = 2$ chiral SUGRA}

The construction of $N = 2$ chiral SUGRA 
starts from the flat, globally supersymmetric spin-(3/2, 1) 
matter multiplet of the chiral theory: 
The flat-space chiral Lagrangian is given by 
\be
L^{0 (+)}_M = - \epsilon^{\mu \nu \rho \sigma} 
              \overline{\tilde \varphi}_{R \mu} \gamma_\rho 
              \partial_\sigma \varphi_{R \nu} 
              - {1 \over 4} (F_{\mu \nu})^2, 
\label{L0+M}
\ee
which is invariant under the right-handed SUSY transformations 
generated by the constant Majorana spinor $\alpha$, 
\ba
\A \A \delta_R A_{\mu} = \sqrt{2} \ \overline \alpha_L 
                         \varphi_{R \mu}, 
\nonu
\A \A \delta_R \varphi_{R \mu} = 0, 
\nonu
\A \A \delta_R \tilde \varphi_{L \mu} 
      = - \sqrt{2} \ i F^{(-)}_{\mu \nu} \gamma^{\nu} \alpha_R, 
\label{RSUSY-M}
\ea
and also under the left-handed SUSY transformations generated 
by another constant Majorana spinor $\tilde \alpha$, 
\ba
\A \A \delta_L A_{\mu} = \sqrt{2} \ \overline{\tilde \alpha}_R 
                         \tilde \varphi_{L \mu}, \nonu
\A \A \delta_L \varphi_{R \mu} = - \sqrt{2} \ i F^{(+)}_{\mu \nu} 
      \gamma^{\nu} \tilde \alpha_L, \nonu
\A \A \delta_L \tilde \varphi_{L \mu} = 0, 
\label{LSUSY-M}
\ea
where $F^{(+)}_{\mu \nu}$ and $F^{(-)}_{\mu \nu}$ represent 
the self-dual and antiself-dual part of $F_{\mu \nu}$, 
respectively. 

If we make the SUSY transformation parameters $\alpha$ 
and $\tilde \alpha$ in (\ref{RSUSY-M}) and (\ref{LSUSY-M}) 
space-time dependent, then terms proportional to 
$(\partial_{\mu} \alpha_R)$ 
and $(\partial_{\mu} \tilde \alpha_L)$ appear 
in $\delta_R L^{0 (+)}_M$ and $\delta_L L^{0 (+)}_M$
respectively. These terms can be eliminated by using 
the transformations $\delta_R \psi_{R \mu} 
= (2/\kappa) \partial_{\mu} \alpha_R$ and 
$\delta_L \tilde \psi_L = (2/\kappa) \partial_{\mu} 
\tilde \alpha_L$ at lowest order in $\kappa$, 
if we add to (\ref{L0+M}) the coupling term of 
$(\psi_{R \mu}, \tilde \psi_{L \mu})$ 
with the Noether current associated with (\ref{L0+M}), 
(\ref{RSUSY-M}) and (\ref{LSUSY-M}): 
Namely, inclusion of the Noether term 
\be
L_{{\rm Noether}} = 
- {\kappa \over 2} (\overline \psi_{L \mu} J_R^{\mu} 
+ \overline{\tilde \psi}_{R \mu} \tilde J_L^{\nu}) 
\label{Noether}
\ee
with 
\ba
\A \A J_R^{\mu} = - 2 \sqrt{2} \ F^{(-) \mu \nu} \varphi_{R \nu}, 
\nonu
\A \A \tilde J_L^{\mu} = - 2 \sqrt{2} \ F^{(+) \mu \nu} 
                         \tilde \varphi_{L \nu}, 
\ea
recovers the right- and left-handed SUSY 
invariance at order $\kappa^0$. Note that 
$J_R^{\mu}$ and $\tilde J_L^{\mu}$ are conserved 
because of the identity 
\ba
\A \A \partial_{[\mu} \varphi_{R \nu]} 
      + {i \over 2} \epsilon_{\mu \nu \rho \sigma} 
      \partial^{\rho} \varphi_R^{\sigma} 
      = {1 \over 2} \gamma_{\alpha} S_{\mu \nu} 
      (\epsilon^{\alpha \beta \rho \sigma} 
      \gamma_{\beta} \partial_{\rho} \varphi_{R \sigma}), 
\nonu
\A \A \partial_{[\mu} \tilde \varphi_{L \nu]} 
      - {i \over 2} \epsilon_{\mu \nu \rho \sigma} 
      \partial^{\rho} \tilde \varphi_L^{\sigma} 
      = - {1 \over 2} \gamma_{\alpha} S_{\mu \nu} 
      (\epsilon^{\alpha \beta \rho \sigma} 
      \gamma_{\beta} \partial_{\rho} 
      \tilde \varphi_{L \sigma}). 
\ea

Furthermore in order to recover the SUSY invariance 
at increasing order in $\kappa$, we start with 
the Lagrangian density 
\ba
\hat{\cal L}^{(+)}_M 
\A = \A - e \ \epsilon^{\mu \nu \rho \sigma} 
        \overline{\tilde \varphi}_{R \mu} \gamma_\rho 
        D^{(+)}_\sigma \varphi_{R \nu} 
      - {e \over 4} (F_{\mu \nu})^2 
\nonu
\A \A + \sqrt{2} \kappa \ e (F^{(-) \mu \nu} 
        \overline \psi_{L \mu} \varphi_{R \nu}
        + F^{(+) \mu \nu} \overline{\tilde \psi}_{R \mu} 
        \tilde \varphi_{L \nu}) 
\label{L1+M}
\ea
obtained by combining the covariantized form of (\ref{L0+M}) 
with the Noether term (\ref{Noether}), 
and examine the variation $\delta \hat{\cal L}^{(+)} = 
\delta ({\cal L}^{(+)}_{SG} + \hat{\cal L}^{(+)}_M)$ 
by using (\ref{RSUSY-SG}), (\ref{LSUSY-SG}), (\ref{RSUSY-M}) 
and (\ref{LSUSY-M}). At order $\kappa$, several terms 
appear in $\delta \hat{\cal L}^{(+)}$. 
The variation of the Maxwell 
action by using $\delta_{R, L} e$, and the variation 
of the Noether term by using $\delta_R \tilde \varphi_L$ 
and $\delta_L \varphi_R$ lead to terms of the form, 
$F^2 \tilde \psi_L$ and $F^2 \psi_R$: 
The explicit calculation shows 
\ba
\A \A \delta_R \hat{\cal L}^{(+)}[F^2 \tilde \psi_L] = 
      {\kappa \over 4} e \ \epsilon^{\mu \nu \rho \sigma} 
      F_{\lambda \nu} F_{\rho \sigma} 
      (\overline{\tilde \psi}_{R \mu} \gamma^{\lambda} \alpha_R 
      - \overline{\tilde \psi}_R^{\lambda} \gamma_{\mu} \alpha_R), 
\nonu
\A \A \delta_L \hat{\cal L}^{(+)}[F^2 \psi_R] = 
      - {\kappa \over 4} e \ \epsilon^{\mu \nu \rho \sigma} 
      F_{\lambda \nu} F_{\rho \sigma} 
      (\overline \psi_{L \mu} \gamma^{\lambda} \tilde \alpha_L 
      - \overline \psi_L^{\lambda} \gamma_{\mu} \tilde \alpha_L), 
\ea
each of which vanishes, however, by means of the identity 
\footnote{\ 
This identity can be proved by multiplying 
$F_{\kappa \nu} F_{\rho \sigma}$ with the identity 
$\delta_{\lambda}^{\kappa} \epsilon^{\mu \nu \rho \sigma} 
= \delta_{\lambda}^{\mu} \epsilon^{\kappa \nu \rho \sigma} 
+ \delta_{\lambda}^{\nu} \epsilon^{\mu \kappa \rho \sigma} 
+ \delta_{\lambda}^{\rho} \epsilon^{\mu \nu \kappa \sigma} 
+ \delta_{\lambda}^{\sigma} \epsilon^{\mu \nu \rho \kappa}$ 
as stated in \cite{FGSN}.}
\be
\epsilon^{\mu \nu \rho \sigma} F_{\lambda \nu} F_{\rho \sigma} 
= {1 \over 4} \delta_{\lambda}^{\mu} 
\epsilon^{\kappa \nu \rho \sigma} F_{\kappa \nu} F_{\rho \sigma}. 
\ee
This cancellation is the same as in the usual SUGRA \cite{FN}. 
Other terms at order $\kappa$ in $\delta \hat{\cal L}^{(+)}$ 
come from the variation of the spin-3/2 action 
in (\ref{L1+M}) by using $\delta_{R, L} e$ and from the variation 
of the Noether terms by using $\delta_{R, L} A$ 
as well as from the first-order variation of 
the self-dual connection in (\ref{RSUSY-SG}) 
and (\ref{LSUSY-SG}): 
Namely, at first order in $\kappa$, we have 
\ba
\delta_R \hat{\cal L}^{(+)} 
\A = \A - i \kappa \ \epsilon^{\mu \nu \rho \sigma} 
        (\overline \alpha_L \gamma^i \tilde \psi_{L \rho}) 
        \overline{\tilde \varphi}_{R \mu} \gamma_i 
        \partial_{\sigma} \varphi_{R \nu} 
\nonu
\A \A + 2 \kappa \left\{ \partial^{[\mu} (\overline \alpha_L 
        \varphi_R^{\nu]}) 
        + {i \over 2} \epsilon^{\mu \nu \rho \sigma}
        \partial_{\rho} (\overline \alpha_L 
        \varphi_{R \sigma}) \right\} 
        \overline \psi_{L \mu} \varphi_{R \nu} 
\nonu
\A \A + 2 \kappa \left\{ \partial^{[\mu} (\overline \alpha_L 
        \varphi_R^{\nu]}) 
        - {i \over 2} \epsilon^{\mu \nu \rho \sigma}
        \partial_{\rho} (\overline \alpha_L 
        \varphi_{R \sigma}) \right\} 
        \overline{\tilde \psi}_{R \mu} \tilde \varphi_{L \nu} 
\label{delR-L+}
\ea
for the right-handed transformations, and 
\ba
\delta_L \hat{\cal L}^{(+)} 
\A = \A - i \kappa \ \epsilon^{\mu \nu \rho \sigma} 
        (\overline{\tilde \alpha}_R \gamma^i \psi_{R \rho}) 
        \overline{\tilde \varphi}_{R \mu} \gamma_i 
        \partial_{\sigma} \varphi_{R \nu} 
      + i \kappa \ \epsilon^{\mu \nu \rho \sigma} 
        (\overline{\tilde \alpha}_R \gamma^i \partial_{\rho} 
        \psi_{R \sigma}) 
        \overline{\tilde \varphi}_{R \mu} 
        \gamma_i \varphi_{R \nu} 
\nonu
\A \A + 2 \kappa \left\{ \partial^{[\mu} 
        (\overline{\tilde \alpha}_R \tilde \varphi_L^{\nu]}) 
        + {i \over 2} \epsilon^{\mu \nu \rho \sigma}
        \partial_{\rho} (\overline{\tilde \alpha}_R 
        \tilde \varphi_{L \sigma}) \right\} 
        \overline \psi_{L \mu} \varphi_{R \nu} 
\nonu
\A \A + 2 \kappa \left\{ \partial^{[\mu} 
        (\overline{\tilde \alpha}_R \tilde \varphi_L^{\nu]}) 
        - {i \over 2} \epsilon^{\mu \nu \rho \sigma} 
        \partial_{\rho} (\overline{\tilde \alpha}_R 
        \tilde \varphi_{L \sigma}) \right\} 
        \overline{\tilde \psi}_{R \mu} \tilde \varphi_{L \nu} 
\label{delL-L+}
\ea
for the left-handed transformations. 
Note that the second term of (\ref{delL-L+}), 
which has no counterpart in (\ref{delR-L+}), 
originates from $\delta \hat{\cal L}^{(+)}_M$ 
for $\delta_L A^{(+)}$ and from $\delta {\cal L}^{(+)}_{SG}$ 
for $\delta_L A^{(+)}$ and $\delta_L \tilde \psi_L$. 
Here we have defined $\delta_L \tilde \psi_L$ 
by using the $A^{(-)}_{ij \mu}$ which is the sum of 
the antiself-dual part of the Ricci rotation coefficients 
$A_{ij \mu}(e)$ and that of $K_{ij \mu}$ given by 
\ba
K_{ij \mu} 
\A = \A {i \over 2} \kappa^2 \{
        (\overline{\tilde \psi}_{R [i} 
        \gamma_{\mid \mu \mid} \psi_{R j]} 
        + \overline{\tilde \psi}_{R [i} 
        \gamma_{\mid j \mid} \psi_{R \mu]} 
        - \overline{\tilde \psi}_{R [j} 
        \gamma_{\mid i \mid} \psi_{R \mu]}) 
\nonu
\A \A + (\overline{\tilde \varphi}_{R [i} 
        \gamma_{\mid \mu \mid} \varphi_{R j]} 
        + \overline{\tilde \varphi}_{R [i} 
        \gamma_{\mid j \mid} \varphi_{R \mu]} 
        - \overline{\tilde \varphi}_{R [j} 
        \gamma_{\mid i \mid} \varphi_{R \mu]}) \}. 
\label{K-N2}
\ea
Namely, the $A^{(-)}_{ij \mu}$ in $\delta_L \tilde \psi_L$ 
is now supposed to be obtained from the ``unphysical" 
Lagrangian density ${\cal L}^{(-)}_{SG}$ which involves 
the pair $(\varphi_{\mu}, \tilde \varphi_{\mu})$ besides 
$(\psi_{\mu}, \tilde \psi_{\mu})$. 

In order to eliminate the terms of (\ref{delR-L+}) and 
(\ref{delL-L+}) we must modify both $\hat{\cal L}^{(+)}_M$ 
and the SUSY transformation laws (\ref{RSUSY-M}) 
and (\ref{LSUSY-M}). As for the terms in (\ref{delR-L+}), 
those proportional to $(\partial_{\mu} \overline \alpha_L)$ 
can be eliminated by adding to $\hat{\cal L}^{(+)}_M$ 
the four-fermion contact terms 
\ba
{\cal L}_{(R)4} 
\A = \A - {\kappa^2 \over 2} e 
        \left( \overline \psi_L^{[\mu} \varphi_R^{\nu]} 
        + {i \over 2} \epsilon^{\mu \nu \rho \sigma} 
        \overline \psi_{L \rho} \varphi_{R \sigma} \right) 
        \overline \psi_{L \mu} \varphi_{R \nu} 
\nonu
\A \A - \kappa^2 e 
        \left( \overline \psi_L^{[\mu} \varphi_R^{\nu]} 
        - {i \over 2} \epsilon^{\mu \nu \rho \sigma} 
        \overline \psi_{L \rho} \varphi_{R \sigma} \right) 
        \overline{\tilde \psi}_{R \mu} \tilde \varphi_{L \nu}, 
\label{LR4}
\ea
while those proportional to 
$(\partial_{\mu} \varphi_{R \nu})$ can be eliminated by adding 
to $\delta_R \tilde \varphi_{L \mu}$ of (\ref{RSUSY-M}) 
the following terms 
\be
\delta'_R \tilde \varphi_{L \mu} 
= i \kappa \left\{
  \left( \overline \psi_{L [\mu} \varphi_{R \nu]} 
  + {i \over 2} \epsilon^{\mu \nu \rho \sigma} 
  \overline \psi_L^{\rho} \varphi_R^{\sigma} \right)
+ \left( \overline{\tilde \psi}_{R [\mu} 
  \tilde \varphi_{L \nu]} 
  + {i \over 2} \epsilon^{\mu \nu \rho \sigma} 
  \overline{\tilde \psi}_R^{\rho} 
  \tilde \varphi_L^{\sigma} \right) \right\} 
  \gamma^{\nu} \alpha_R. 
\ee

As for the terms in (\ref{delL-L+}), on the other hand, 
we first note that the sum of the first 
and second terms in (\ref{delL-L+}) become 
$- 2i \kappa \ \epsilon^{\mu \nu \rho \sigma} \partial_{\rho} 
(\overline{\tilde \alpha}_R \tilde \varphi_{L \sigma}) 
\overline \psi_{L \mu} \varphi_{R \nu}$ after Fierz 
transformations. Then the terms proportional to 
$(\partial_{\mu} \overline{\tilde \alpha}_R)$ 
can be eliminated by adding to $\hat{\cal L}^{(+)}_M$ 
the four-fermion contact terms 
\ba
{\cal L}_{(L)4} 
\A = \A - \kappa^2 e 
        \left( \overline{\tilde \psi}_R^{[\mu} 
        \tilde \varphi_L^{\nu]} 
        - {i \over 2} \epsilon^{\mu \nu \rho \sigma} 
        \overline{\tilde \psi}_{R \rho} 
        \tilde \varphi_{L \sigma} \right) 
        \overline \psi_{L \mu} \varphi_{R \nu} 
\nonu
\A \A - {\kappa^2 \over 2} e 
      \left( \overline{\tilde \psi}_R^{[\mu} 
      \tilde \varphi_L^{\nu]} 
      - {i \over 2} \epsilon^{\mu \nu \rho \sigma} 
      \overline{\tilde \psi}_{R \rho} 
      \tilde \varphi_{L \sigma} \right) 
      \overline{\tilde \psi}_{R \mu} \tilde \varphi_{L \nu}, 
\label{LL4}
\ea
while the remaining terms of (\ref{delL-L+}) 
proportional to $(\partial_{\mu} \tilde \varphi_{L \nu})$ 
can be eliminated by adding to 
$\delta_L \varphi_{R \mu}$ in (\ref{LSUSY-M}) 
the following terms 
\be
\delta'_L \varphi_{R \mu} 
= i \kappa \left\{
  \left( \overline \psi_{L [\mu} \varphi_{R \nu]} 
  - {i \over 2} \epsilon^{\mu \nu \rho \sigma} 
  \overline \psi_L^{\rho} \varphi_R^{\sigma} \right)
+ \left( \overline{\tilde \psi}_{R [\mu} 
  \tilde \varphi_{L \nu]} 
  - {i \over 2} \epsilon^{\mu \nu \rho \sigma} 
  \overline{\tilde \psi}_R^{\rho} 
  \tilde \varphi_L^{\sigma} \right) \right\} 
  \gamma^{\nu} \tilde \alpha_L. 
\label{dL-phi}
\ee
Here we also note that all terms of (\ref{LR4}) through 
(\ref{dL-phi}) contain the (anti)self-dual part of 
$\overline \psi_{L [\mu} \varphi_{R \nu]}$ or 
$\overline{\tilde \psi}_{R [\mu} \tilde \varphi_{L \nu]}$. 

To summarize, in order to recover the right- and left-handed 
SUSY invariance at order $\kappa$, 
we have modified the SUSY transformations 
of $\delta_R \tilde \varphi_{L \mu}$ 
and $\delta_L \varphi_{R \mu}$ in (\ref{RSUSY-M}) and 
(\ref{LSUSY-M}) as 
\ba
\A \A \delta_R \tilde \varphi_{L \mu} 
      = - \sqrt{2} \ i \hat F^{(-)}_{\mu \nu} 
      \gamma^{\nu} \alpha_R, 
\nonu
\A \A \delta_L \varphi_{R \mu} 
      = - \sqrt{2} \ i \hat F^{(+)}_{\mu \nu} 
      \gamma^{\nu} \tilde \alpha_L, 
\label{Mod-phi}
\ea
where $\hat F^{(\pm)}_{\mu \nu}$ represent the self-dual 
and antiself-dual part of $\hat F_{\mu \nu}$ defined by 
\be
\hat F_{\mu \nu} := F_{\mu \nu} - \sqrt{2} \kappa 
(\overline \psi_{L [\mu} \varphi_{R \nu]} 
+ \overline{\tilde \psi}_{R [\mu} \tilde \varphi_{L \nu]}). 
\label{F-hat}
\ee
The $\hat{\cal L}^{(+)}_M$ of (\ref{L1+M}) has also been 
modified as 
\footnote{\ 
Note that since the second term of ${\cal L}_{(R)4}$ 
coincides with the first term of ${\cal L}_{(L)4}$, 
we take only one of them.}
\ba
{\cal L}^{(+)}_M 
\A = \A - e \ \epsilon^{\mu \nu \rho \sigma} 
        \overline{\tilde \varphi}_{R \mu} \gamma_\rho 
        D^{(+)}_\sigma \varphi_{R \nu} 
      - {e \over 4} (F_{\mu \nu})^2 
\nonu
\A \A + {\kappa \over \sqrt{2}} e \ \{ 
        (F^{(-) \mu \nu} + \hat F^{(-) \mu \nu}) 
        \overline \psi_{L \mu} \varphi_{R \nu}
        + (F^{(+) \mu \nu} + \hat F^{(+) \mu \nu}) 
        \overline{\tilde \psi}_{R \mu} 
        \tilde \varphi_{L \nu}) \} 
\nonu
\A \A + {i \over 2} \kappa^2 e \ \epsilon^{\mu \nu \rho \sigma} 
        (\overline \psi_{L \rho} \varphi_{R \sigma}) 
        \overline{\tilde \psi}_{R \mu} \tilde \varphi_{L \nu}. 
\label{L2+M}
\ea
The last term of (\ref{L2+M}) is needed to cancel 
those terms in the variation of the Noether term 
with respect to $\delta_{R,L} A$ at order $\kappa$, 
and it plays an important role in solving 
the consistency problem as will be explained later. 

At order $\kappa^2$, although the transformations 
$\delta_{R, L} A^{(+)}_{ij \mu}$ should be corrected 
to recover the SUSY invariance within the first-order 
formalism, this task will be complicated \cite{FS}. 
Therefore, we turn to the second-order formalism 
in order to minimize complication, as was done 
in constructing the usual $N = 3$ SUGRA \cite{Free}: 
Namely, the equation for $A^{(+)}_{ij \mu}$ derived from 
${\cal L}^{(+)}_{N = 2} = {\cal L}^{(+)}_{SG} 
+ {\cal L}^{(+)}_M$ is solved as 
\be
A^{(+)}_{ij \mu} = A^{(+)}_{ij \mu}(e) + K^{(+)}_{ij \mu}, 
\label{solA+}
\ee
where $A^{(+)}_{ij \mu}(e)$ is the self-dual part 
of the Ricci rotation coefficients $A_{ij \mu}(e)$, 
while $K^{(+)}_{ij \mu}$ is that of $K_{ij \mu}$ given by 
(\ref{K-N2}). If we use the solution (\ref{solA+}), 
then all the terms of order $\kappa^2$, some of which 
are derived from the terms added in (\ref{Mod-phi}) 
and (\ref{L2+M}), cancel identically. 
Finally at order $\kappa^3$, quintic terms 
of spin-3/2 fields appear but they indeed 
cancel among themselves.

\newsection{The manifest $O(2)$ invariance}

Let us show that the resultant $N = 2$ chiral Lagrangian 
has a manifest $O(2)$ invariance which rotates 
the two pairs of spin-3/2 fields into one another as 
in the usual $N = 2$ SUGRA \cite{FN,Nieu}. 
In fact, we can rewrite the ${\cal L}^{(+)}_{N = 2} 
= {\cal L}^{(+)}_{SG} + {\cal L}^{(+)}_M$ as 
\ba
{\cal L}^{(+)}_{N = 2} 
\A = \A - {i \over {2 \kappa^2}} 
        e \ \epsilon^{\mu \nu \rho \sigma} 
        e^i_{\mu} e^j_{\nu} R^{(+)}_{ij \rho \sigma} 
      - e \ \epsilon^{\mu \nu \rho \sigma} 
        \overline{\tilde \psi}^I_{R \mu} \gamma_\rho 
        D^{(+)}_\sigma \psi^I_{R \nu} 
      - {e \over 4} (F_{\mu \nu})^2 
\nonu
\A \A + {\kappa \over{2 \sqrt{2}}} e \ \{ 
        (F^{(-) \mu \nu} + \hat F^{(-) \mu \nu}) 
        \overline \psi^I_{L \mu} \psi^J_{R \nu} 
        + (F^{(+) \mu \nu} + \hat F^{(+) \mu \nu}) 
        \overline{\tilde \psi}^I_{R \mu} 
        \tilde \psi^J_{L \nu} \} \epsilon^{IJ} 
\nonu
\A \A + {i \over 8} \kappa^2 
        e \ \epsilon^{\mu \nu \rho \sigma} 
        (\overline \psi^I_{L \mu} \psi^J_{R \nu}) 
        \overline{\tilde \psi}^K_{R \rho} 
        \tilde \psi^L_{L \sigma} 
        \epsilon^{IJ} \epsilon^{KL}, 
\label{LN2+}
\ea
where we denote the pairs by $(\psi^I_{R \mu}, 
\tilde \psi^I_{L \mu})$ $(I, J, \dots = 1, 2)$ 
\footnote{\ 
A summation convention for the internal-symmetry indices 
$I, J, \dots$ is used and $\epsilon^{IJ} = - \epsilon^{JI}$.}
instead of $(\psi_{R \mu}, \tilde \psi_{L \mu})$ 
and $(\varphi_{R \mu}, \tilde \varphi_{L \mu})$, 
and accordingly the definition of $\hat F_{\mu \nu}$, 
(\ref{F-hat}), is expressed by 
\be
\hat F_{\mu \nu} := F_{\mu \nu} - {\kappa \over \sqrt{2}} 
(\overline \psi^I_{L [\mu} \psi^J_{R \nu]} 
+ \overline{\tilde \psi}^I_{R [\mu} \tilde \psi^J_{L \nu]})
\epsilon^{IJ}. 
\ee
The chiral Lagrangian density of (\ref{LN2+}) is invariant 
under the right-handed SUSY transformations generated by 
$\alpha^I$ 
\ba
\A \A \delta_R e^i_{\mu} 
               = i \kappa \ \overline \alpha^I_L 
               \gamma^i \tilde \psi^I_{L \mu}, 
               \ \ \ \ \ 
      \delta_R A_{\mu} 
               = \sqrt{2} \ \epsilon^{IJ} 
               \overline \alpha^I_L \psi^J_{R \mu}, 
\nonu
\A \A \delta_R \psi^I_{R \mu} 
               = {2 \over \kappa} D^{(+)}_{\mu} \alpha^I_R, 
\nonu
\A \A \delta_R \tilde \psi^I_{L \mu} 
               = \sqrt{2} \ i \ \epsilon^{IJ} 
               \hat F^{(-)}_{\mu \nu} \gamma^{\nu} \alpha^J_R, 
\label{RSUSY-N2}
\ea
and also under the left-handed SUSY transformations 
generated by $\tilde \alpha^I$ 
\ba
\A \A \delta_L e^i_{\mu} 
               = i \kappa \ \overline{\tilde \alpha}^I_R 
               \gamma^i \psi^I_{R \mu}, 
               \ \ \ \ \ 
      \delta_L A_{\mu} 
               = \sqrt{2} \ \epsilon^{IJ} 
               \overline{\tilde \alpha}^I_R 
               \tilde \psi^J_{L \mu}, 
\nonu
\A \A \delta_L \tilde \psi^I_{L \mu} 
               = {2 \over \kappa} D^{(-)}_{\mu} 
               \tilde \alpha^I_L, 
\nonu
\A \A \delta_L \psi^I_{R \mu} 
      = \sqrt{2} \ i \ \epsilon^{IJ} 
      \hat F^{(+)}_{\mu \nu} \gamma^{\nu} \tilde \alpha^J_L. 
\label{LSUSY-N2}
\ea

Within the first-order formalism, the $A^{(+)}_{ij \mu}$ 
appears in the chiral Lagrangian density 
and it is taken as one of the independent variables. 
As for the $A^{(-)}_{ij \mu}$, which is used 
in the left-handed SUSY transformations of (\ref{LSUSY-N2}), 
however, we define it to be derived from 
the ``unphysical" Lagrangian density (\ref{LSG-}) 
with $(\psi_{R \mu}, \tilde \psi_{L \mu})$ 
being replaced by $(\psi^I_{R \mu}, \tilde \psi^I_{L \mu})$: 
Namely, the $A^{(-)}_{ij \mu}$ is fixed as 
the sum of the antiself-dual part of the Ricci rotation 
coefficients $A_{ij \mu}(e)$ and that of $K_{ij \mu}$ 
of (\ref{K-N2}), which now reads 
\be
K_{ij \mu} = {i \over 2} \kappa^2 
             (\overline{\tilde \psi}^I_{R [i} 
             \gamma_{\mid \mu \mid} \psi^I_{R j]} 
             + \overline{\tilde \psi}^I_{R [i} 
             \gamma_{\mid j \mid} \psi^I_{R \mu]} 
             - \overline{\tilde \psi}^I_{R [j} 
             \gamma_{\mid i \mid} \psi^I_{R \mu]}). 
\ee
When the reality condition, 
\be
\overline{e^i_{\mu}} = e^i_{\mu} {\rm\ \ and\ \ } 
\tilde \psi^I_{\mu} = \psi^I_{\mu}, 
\label{reality}
\ee
are imposed, 
\footnote{
The bar of $e^i_{\mu}$ in (\ref{reality}) means 
the complex conjugate. }
the $A^{(-)}_{ij \mu}$, which is used in (\ref{LSUSY-N2}), 
becomes just the complex conjugate 
of the $A^{(+)}_{ij \mu}$ of (\ref{solA+}).

\newsection{Compatibility with the reality condition}

We now discuss the consistency problem and the relation to 
the usual $N = 2$ SUGRA within the second-order formalism. 
If we take the self-dual connection 
$A^{(+)}_{ij \mu}$ in ${\cal L}^{(+)}_{N = 2}$ as the function 
of the tetrad and spin-3/2 fields by solving the equation 
$\delta {\cal L}^{(+)}_{N = 2}/\delta A^{(+)} = 0$, 
then the torsion part of ${\cal L}^{(+)}_{N = 2}$ 
involves a four-fermion contact term 
\be
{i \over 8} e \ \epsilon^{\mu \nu \rho \sigma} 
T_{\lambda \mu \nu} T{^{\lambda}}_{\rho \sigma} = 
- {i \over{16}} \kappa^2 e \ \epsilon^{\mu \nu \rho \sigma} 
(\overline{\tilde \psi}^I_{R \mu} \gamma_{\lambda} 
\psi^K_{R \nu}) 
\overline{\tilde \psi}^J_{R \rho} \gamma^{\lambda} 
\psi^L_{R \sigma} \epsilon^{IJ} \epsilon^{KL}, 
\label{4-Fermi}
\ee
since the torsion tensor is related to $K_{\lambda \mu \nu}$ 
by $T_{\lambda \mu \nu} = 2 K_{\lambda [\mu \nu]}$. 
When the reality condition (\ref{reality}) is satisfied, 
the right-hand side of (\ref{4-Fermi}) becomes 
pure imaginary, indicating that if such a term really 
survives in ${\cal L}^{(+)}_{N = 2}$, it will give 
additional equations for spin-3/2 fields which cause 
inconsistency \cite{TS1}. 
However, since the last term in 
${\cal L}^{(+)}_{N = 2}$ can be rewritten as 
\be
{i \over 8} \kappa^2 e \ \epsilon^{\mu \nu \rho \sigma} 
(\overline \psi^I_{L \mu} \psi^J_{R \nu}) 
\overline{\tilde \psi}^K_{R \rho} \tilde \psi^L_{L \sigma} 
\epsilon^{IJ} \epsilon^{KL} 
= {i \over{16}} \kappa^2 e \ \epsilon^{\mu \nu \rho \sigma} 
(\overline{\tilde \psi}^I_{R \mu} \gamma_{\lambda} 
\psi^K_{R \nu}) 
\overline{\tilde \psi}^J_{R \rho} \gamma^{\lambda} 
\psi^L_{R \sigma} \epsilon^{IJ} \epsilon^{KL} 
\ee
by using a Fierz transformation, this exactly cancels with 
the term of (\ref{4-Fermi}), and therefore any inconsistency 
of field equations does not arise at the classical level 
even if the reality condition is imposed. 
In other words, the SUSY invariance of 
${\cal L}^{(+)}_{N = 2}$ solves the consistency problem 
which would arise without the SUSY invariance. 
Indeed, the ${\cal L}^{(+)}_{N = 2}$ of $N = 2$ chiral SUGRA 
with the reality condition (\ref{reality}) is reduced to 
that of the usual one up to an imaginary boundary term: 
Namely, we have 
\be
{\cal L}^{(+)}_{N = 2} [A^{(+)}(e, \psi^I)] 
= {\cal L}_{N = 2 {\rm\ usual\ SUGRA}} [A(e, \psi^I)] 
- {1 \over 4} \partial_{\mu} 
(e \ \epsilon^{\mu \nu \rho \sigma} 
\overline \psi^I_{\nu} \gamma_{\rho} \psi^I_{\sigma}). 
\ee
This Lagrangian density is invariant under 
the right- and left-handed SUSY transformations 
of (\ref{RSUSY-N2}) and (\ref{LSUSY-N2}), 
which are now complex conjugate of each other.

\newsection{Conclusion and discussion}

In this paper we have constructed $N = 2$ chiral SUGRA 
compatible with the reality condition 
using the SUSY transformation parameters without constraint. 
We have also shown within the second-order formalism 
that the formulation is reduced to the usual $N = 2$ SUGRA 
up to an imaginary boundary term when the reality condition 
is satisfied. 

Ashtekar's canonical formulation of $N = 2$ chiral 
SUGRA derived from (\ref{LN2+}) does not contain 
any auxiliary field: This point is in sharp contrast with 
that of $N = 2$ two-form SUGRA \cite{KS}. 
So detailed comparison of our results with the two-form SUGRA 
is desirable at the level of canonical formulation. 

Construction of $N = 3$ chiral SUGRA will be 
a straightforward task by using the method of this paper, 
except for the following aspect: 
Namely, since the $N = 3$ theory contains 
both spin-1/2 and -3/2 fields, the torsion tensor 
for the real tetrad is fixed as 
\be
T_{\lambda \mu \nu} = - {i \over 2} \kappa^2 
\overline \psi^I_{[\mu} 
\gamma_{\mid \lambda \mid} \psi^I_{\nu]} 
- {\kappa^2 \over 4} \epsilon_{\lambda \mu \nu \sigma} 
\overline \chi \gamma_5 \gamma^{\sigma} \chi 
\ \ (I = 1, 2, 3), 
\label{torsion-N3}
\ee
and therefore we must consider the consistency 
problem for the term 
$(i/8) e \ \epsilon^{\mu \nu \rho \sigma} 
T_{\lambda \mu \nu}$ $\times T{^{\lambda}}_{\rho \sigma}$ 
with the torsion being given by (\ref{torsion-N3}). 
The problem is whether this term is canceled 
with another term required by the SUSY invariance or not. 
The construction of $N = 3$ chiral SUGRA 
and further extensions are now being investigated. 

An attempt at supersymmetric extension of 
loop quantum gravity \cite{RS,Ro} has been made 
for $N = 1$ chiral SUGRA \cite{AU}. We expect that 
its further extension to the theory with $N > 1$ 
can be made, now that $N = 2$ chiral SUGRA 
compatible with the reality condition 
has been constructed.

{\large{\bf{Ackowledgments}}}

We would like to thank Professor Y. Tanii and other members 
of Physics Department at Saitama University 
for discussions and encouragements.




\newpage
\begin{thebibliography}{100}

\bibitem{AA} 
Ashtekar A 
1986 {\it Phys. Rev. Lett.} {\bf 57} 2244; 
1987 {\it Phys. Rev.} D {\bf 36} 1587 

\bibitem{JS} 
Jacobson T and Smolin L 
1987 {\it Phys. Lett.} {\bf 196B} 39; 
1988 {\it Class. Quantum Grav.} {\bf 5} 583 

\bibitem{RS} 
Rovelli C and Smolin L 
1990 {\it Nucl. Phys.} B {\bf 331} 80 

\bibitem{Ro} 
Rovelli C 
1997 Loop Quantum Gravity 
(Review written for the electronic journal 
LIVING REVIEWS) {\it Preprint} gr-qc/9710008 

\bibitem{Do} 
Dolan B P 
1989 {\it Phys. Lett.} {\bf 233B} 89 

\bibitem{TS1} 
Tsuda M, Shirafuji T and Xie H 
1995 {\it Class. Quantum Grav.} {\bf 12} 3067 

\bibitem{TS2} 
Tsuda M and Shirafuji T 
1996 {\it Phys. Rev.} D {\bf 54} 2960 

\bibitem{JJ1} 
Jacobson T 
1988 {\it Class. Quantum Grav.} {\bf 5} L143 

\bibitem{ART} 
Ashtekar A, Romano J D and Tate R S 
1989 {\it Phys. Rev.} D {\bf 40} 2572 

\bibitem{JJ2} 
Jacobson T 
1988 {\it Class. Quantum Grav.} {\bf 5} 923 

\bibitem{CDJ} 
Capovilla R, Dell J, Jacobson T and Mason L 
1991 {\it Class. Quantum Grav.} {\bf 8} 41 

\bibitem{KS} 
Kunitomo H and Sano T 
1993 {\it Prog. Theor. Phys. Suppl.} {\bf 114} 31; 
1993 {\it Int. J. Mod. Phys.} D {\bf 1} 559 

\bibitem{FNF} 
Freedman D Z, van Nieuwenhuizen P and Ferrara S 
1976 {\it Phys. Rev.} D {\bf 13} 3214; 
Freedman D Z and van Nieuwenhuizen P 
1976 {\it Phys. Rev.} D {\bf 14} 912 

\bibitem{DZ} 
Deser S and Zumino B 
1976 {\it Phys. Lett.} {\bf 62B} 335 

\bibitem{TS3} 
Tsuda M and Shirafuji T 
1997 Minimal Off-Shell Version of $N = 1$ Chiral Supergravity 
{\it Preprint} STUPP-97-152, gr-qc/9711077 

\bibitem{FN} 
Ferrara S and van Nieuwenhuizen P 
1976 {\it Phys. Rev. Lett.} {\bf 37} 1669 

\bibitem{FGSN} 
Ferrara S, Gliozzi F, Scherk J and van Nieuwenhuizen P 
1976 {\it Nucl. Phys.} B {\bf 117} 333 

\bibitem{West} 
West P 
1990 {\it Introduction to Supersymmetry and Supergravity} 
(Singapore: World Scientific) 

\bibitem{Nieu} 
van Nieuwenhuisen P 
1981 {\it Phys. Rep.} {\bf 68} 189 

\bibitem{FS} 
Freedman D Z and Schwarz J H 
1977 {\it Phys. Rev.} D {\bf 15} 1007 

\bibitem{Free} 
Freedman D Z 
1977 {\it Phys. Rev. Lett.} {\bf 38} 105 

\bibitem{AU} 
Armand-Ugon D, Gambini R, Obreg\'on O and Pullin J 
1996 {\it Nucl. Phys.} B {\bf 460} 615 





\end{thebibliography}
\end{document}